\documentclass[12pt]{article}
\usepackage[dvips]{graphicx}
\usepackage{amsthm}
\usepackage{epsfig}  
\usepackage{amssymb}

\def\be{\begin{equation}}
\def\bea{\begin{eqnarray}}
\def\ee{\end{equation}}
\def\eea{\end{eqnarray}}

\def\pt{\partial}

\def\gm{\gamma}

\def\dt{\delta}
\def\Dt{\Delta}
\def\eps{\varepsilon}
\def\ffi{\varphi}

\def\kp{\kappa}
\def\La{\Lambda}
\def\la{\lambda}

\def\Th{\Theta}

\def\Om{\Omega}
\def\om{\omega}

\def\dd{\mbox{d}}

\def\ln{\mbox{ln}}

\def\calH{{\cal H}}
\def\calP{{\cal P}}

\def\calS{{\cal S}}

\def\cn{\mbox{cn}}

\begin{document}

\begin{center}
{\Large \bf Capture into resonance and escape from it in a forced nonlinear pendulum}\\

\vskip 10mm

{\Large A. I. Neishtadt$^{1,2}$, A.~A.~Vasiliev$^{1}$,  A. V. Artemyev$^{1}$,\\

\vskip 5mm

{\large \it $^{1}$ Space Research Institute, Moscow, Russia}\\
{\large \it $^{2}$ Department of Mathematical Sciences, \\ Loughborough University, UK}\\
}
\end{center}

\section*{Abstract}
We study dynamics of a nonlinear pendulum under a periodic force with small amplitude and slowly decreasing frequency. It is well known that when the frequency of the external force passes through the value of the frequency of the unperturbed pendulum's oscillations, the pendulum can be captured into the resonance. The captured pendulum  oscillates in such a way that the resonance is preserved, and the amplitude of the oscillations accordingly grows. We consider this problem in the frames of a standard Hamiltonian approach to resonant phenomena in slow-fast Hamiltonian systems developed earlier, and evaluate the probability of capture into the resonance. If the system passes the resonance at small enough initial amplitudes of the pendulum, the capture occurs with necessity (so-called autoresonance). In general, the probability of capture varies between one and zero, depending on the initial amplitude. We demonstrate that a pendulum captured at small values of its amplitude escapes from the resonance in the domain of oscillations close to the separatrix of the pendulum, and evaluate the amplitude of the oscillations at the escape.

\section{Introduction}

A pendulum under the action of an external force is an important and ubiquitous model in various areas of nonlinear dynamics. One of the interesting and important in applications phenomena is capture of the pendulum's oscillations into resonance with the oscillations of the external force. Consider first the pendulum initially at rest, and the external force of small amplitude $\eps$ and frequency equal to the frequency of the pendulum's linear oscillations $\om_0$. It is known that such a force can increase the pendulum's amplitude up to a value of order $\eps^{1/3}$. At larger amplitudes the dependence of the pendulum's frequency on the amplitude (i.e., the pendulum's nonlinearity) results in breakup of the resonance. However, if the frequency of the external force is not constant but slowly decreasing with time at a rate $\dt$, the so called autoresonance phenomenon occurs (see, e.g., \cite{Friedlandscholar} and the references therein). When the frequency of the force passes through the value $\om_0$, the pendulum is captured into the resonance with the force, and the amplitude of its oscillations increases in such a way that the pendulum stays in the resonance.

This phenomenon has been studied in numerous works starting from the pioneering papers of V. I. Veksler and E. M. McMillan \cite{Veksler45,McMillan45} where a crucial role of the autoresonance in particle accelerators was demonstrated. However, it is methodically important and interesting to consider this phenomenon using the standard and well-developed Hamiltonian approach to resonant phenomena in nonlinear systems \cite{AKN}. We do this in the present paper. We show that automatic capture into resonance (capture with probability one) has place not only for zero initial amplitude of the pendulum, but also for small non-zero initial amplitudes. We also consider the case of larger initial amplitudes and describe the capture into the resonance and evaluate its probability. A captured pendulum escapes from the resonance at a certain final amplitude of oscillations. We describe this phenomenon and find the energy of the pendulum at the escape.

We study mostly the important ``adiabatic'' case, when the variation rate of the external frequency is much smaller than the amplitude of the external force: $\dt \ll \eps$. However, some conclusions are made also for the case when  $\dt \sim \eps$. In particular, we show that capture into the resonance is impossible, if $\eps$ is smaller than a certain threshold that depend on the value of $\dt$. This agrees with so-called threshold phenomenon (see \cite{Friedlandscholar}).

\section{Capture into resonance at small initial values of the amplitude}\label{sec2}

We start at the Hamiltonian of a pendulum under the action of a time-periodic external forcing:
\be
H = \frac{P^2}{2} - \om_0^2 \cos Q + \eps \cos \psi \cdot Q.
\label{0.1}
\ee
Here $(P,Q)$ are canonically conjugate momentum and coordinate, $\om_0$ is a frequency of linear oscillations, $\eps \ll1$ and $\psi$ are the amplitude and the phase of the external forcing respectively. Assume that $\dot \psi = \om(\delta t)$ is a slowly varying frequency of the external forcing, $0 <\delta \ll 1$.

A standard way to study this system near the 1:1 resonance is to introduce the action-angle variables of the unperturbed pendulum as a new pair of canonical variables, average the Hamiltonian near the resonance, and expand it into series. This approach is implemented in Section \ref{sec3} of this paper. However, when the initial amplitude of the pendulum is zero or small, of order $\eps^{1/3}$, it is easier to apply a different method. Namely, one can expand the cosine in (\ref{0.1}) and use symplectic polar coordinates instead of the exact action-angle variables.  Below in this section we use this latter approach. The results obtained by the two methods at small values of the initial amplitude asymptotically agree with each other.

Thus, in this section we study the case of small amplitude of the pendulum's oscillations: $|Q| \ll 1$. Expanding $\cos Q$ into series and omitting a constant, we find in the main approximation
\be
H = \frac{P^2}{2} + \om_0^2 \frac{Q^2}{2}- \om_0^2 \frac{Q^4}{24} + \eps \cos \psi \cdot Q.
\label{0.2}
\ee
Introduce new canonical variables (so-called symplectic polar coordinates) $\rho$ and $\phi$:
\be
Q = \sqrt{2\rho/\om_0} \sin\phi, \,\,\,\,\, P =  \sqrt{2\rho\om_0} \cos\phi.
\label{0.3}
\ee
In terms of the new variables the Hamiltonian takes the form
\be
H = \rho\om_0 - \frac{\rho^2}{6} \sin^4\phi + \eps \cos \psi \cdot \sqrt{2\rho/\om_0} \sin\phi.
\label{0.4}
\ee
We consider the situation when the system is close to the 1:1 resonance, i.e., when $\dot \phi \approx \dot\psi$. Introduce the {\it resonance phase} $\gm = \phi - \psi$ as a new variable. To do this, make a canonical change of variables $(\rho, \phi) \mapsto (\tilde \rho, \gm)$ defined with generating function $W(\tilde \rho,\phi)= \tilde \rho(\phi - \psi)$. In the new variables, the Hamiltonian is $\tilde H = H + \pt W/\pt t = H - \tilde\rho \om$.  One can average  over the fast phase and obtain the Hamiltonian averaged near the resonance (we omit tildes):
\be
H = \rho\om_0 - \frac{\rho^2}{16} + \frac12 \eps \sqrt{2\rho/\om_0} \sin \gm - \rho \om.
\label{0.5}
\ee
Introduce another pair of canonical variables $(x,y)$:
\be
x = \sqrt{2\rho} \sin\gm, \,\,\,\,\, y =  \sqrt{2\rho} \cos\gm.
\label{0.6}
\ee
In these new variables, the Hamiltonian is
\be
H = (\om_0-\om) \frac{x^2+y^2}{2} - \frac{(x^2+y^2)^2}{64} + \frac{\eps}{2\sqrt{\om_0}}x.
\label{0.7}
\ee
Rescaling the Hamiltonian and time: $H \rightarrow -64 H, \, t \rightarrow t/64$, and changing $x \rightarrow -x$ we obtain the Hamiltonian in the standard form
\be
H = (x^2+y^2)^2 - \la (x^2+y^2) + \mu \, x, \,\,\, \la = 32(\om_0-\om),\,\,\, \mu = 32\frac{\eps}{\sqrt{\om_0}}.
\label{0.8}
\ee
This Hamiltonian appears in many resonant problems in celestial mechanics, atomic and plasma physics (see, e.g., \cite{Sincl,Gr,Henrard83,NeishtadtTimofeev:1987,KotStup:1991,NeishtadtVasiliev:2005}). Properties of a system with this Hamiltonian were thoroughly investigated in \cite{Neishtadt:1975}, and here we just put forward some of results obtained in that paper.

In Hamiltonian (\ref{0.8}) parameter $\mu$ is a positive constant, and parameter $\la$ is a slowly varying function of time: $\dot\la \sim \delta$. We assume that $\dot \la >0$, i.e. that the frequency $\om$ is decreasing with time.

\begin{figure}[htbp]
\begin{tabular}{cc}
\vspace*{-10mm} \hspace*{-4mm} \psfig{file=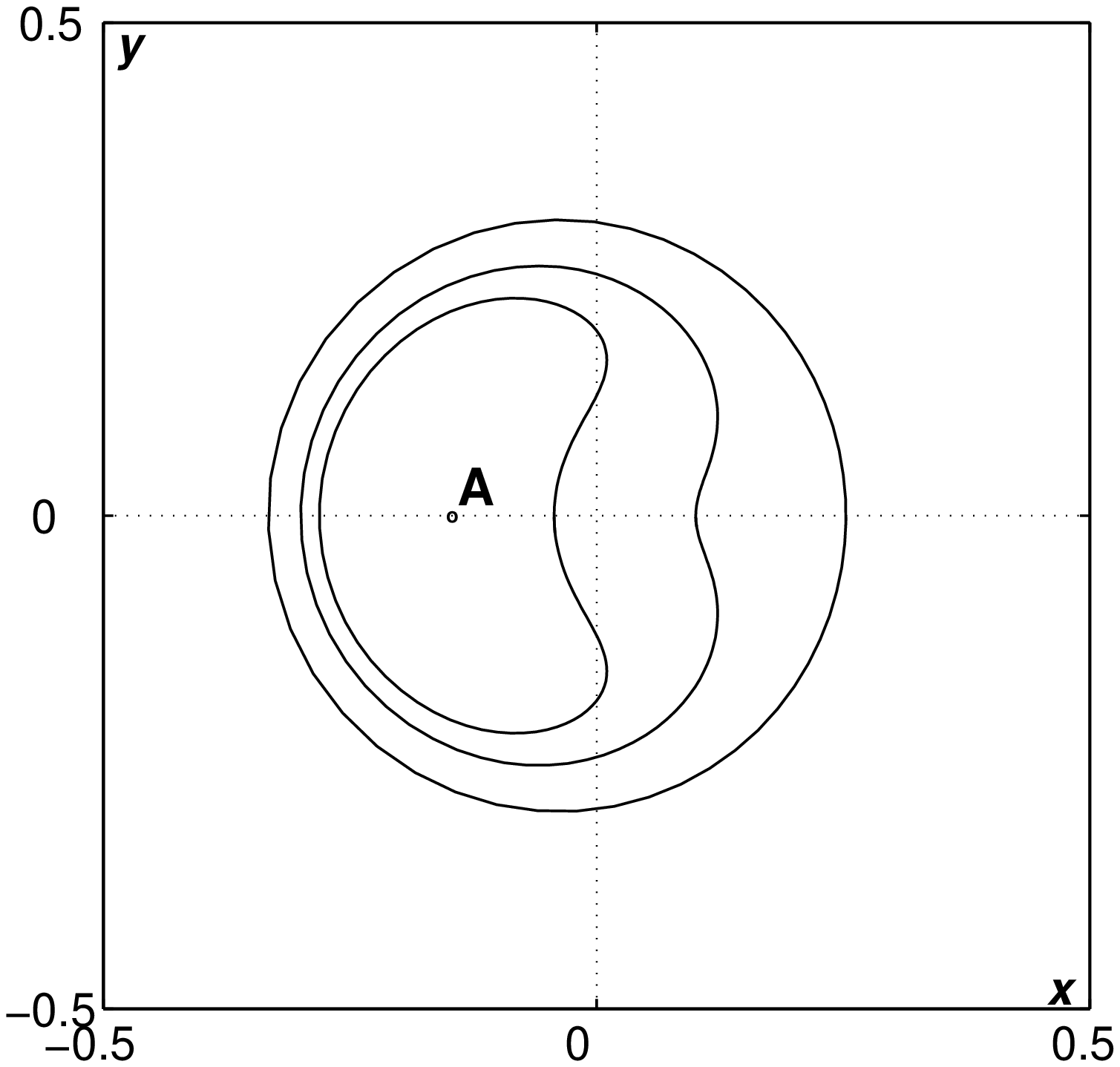,width=200pt} & \hspace*{4mm}
\psfig{file=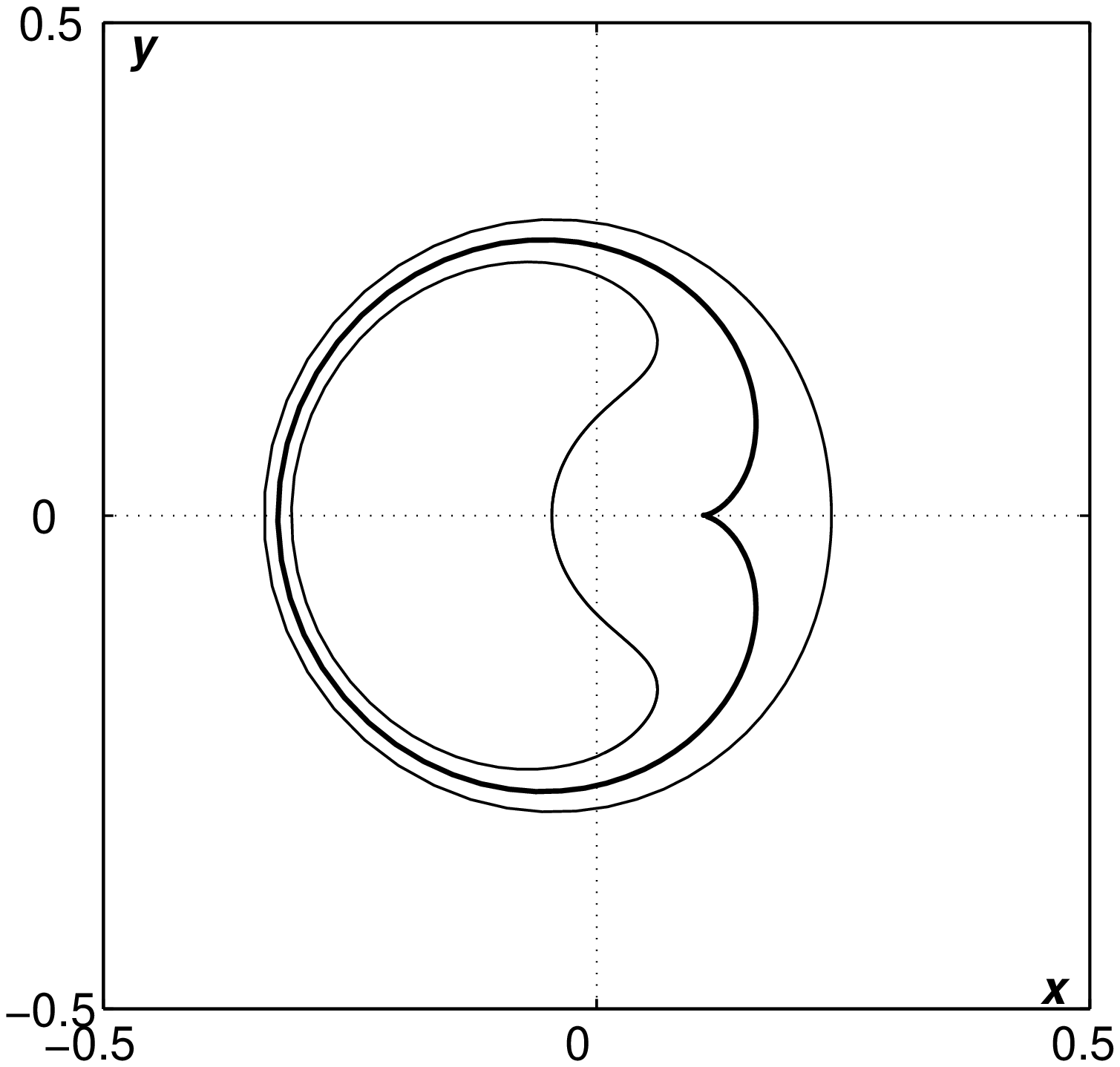,width=200 pt}\vspace*{10mm}\\
\vspace*{0mm} \hspace*{-1mm}(a) &\hspace*{5mm} (b)
\end{tabular}
\caption{\small{Phase portraits of the system at fixed values of
$\la$; $\tilde\mu = 0.01$. a) $\la  = 0.05 < \la_*$, b) $\la =
0.069624 \approx \la_*$.}}\label{portraits12}
\end{figure}

\begin{figure}[h]
\hspace*{2mm} \psfig{file=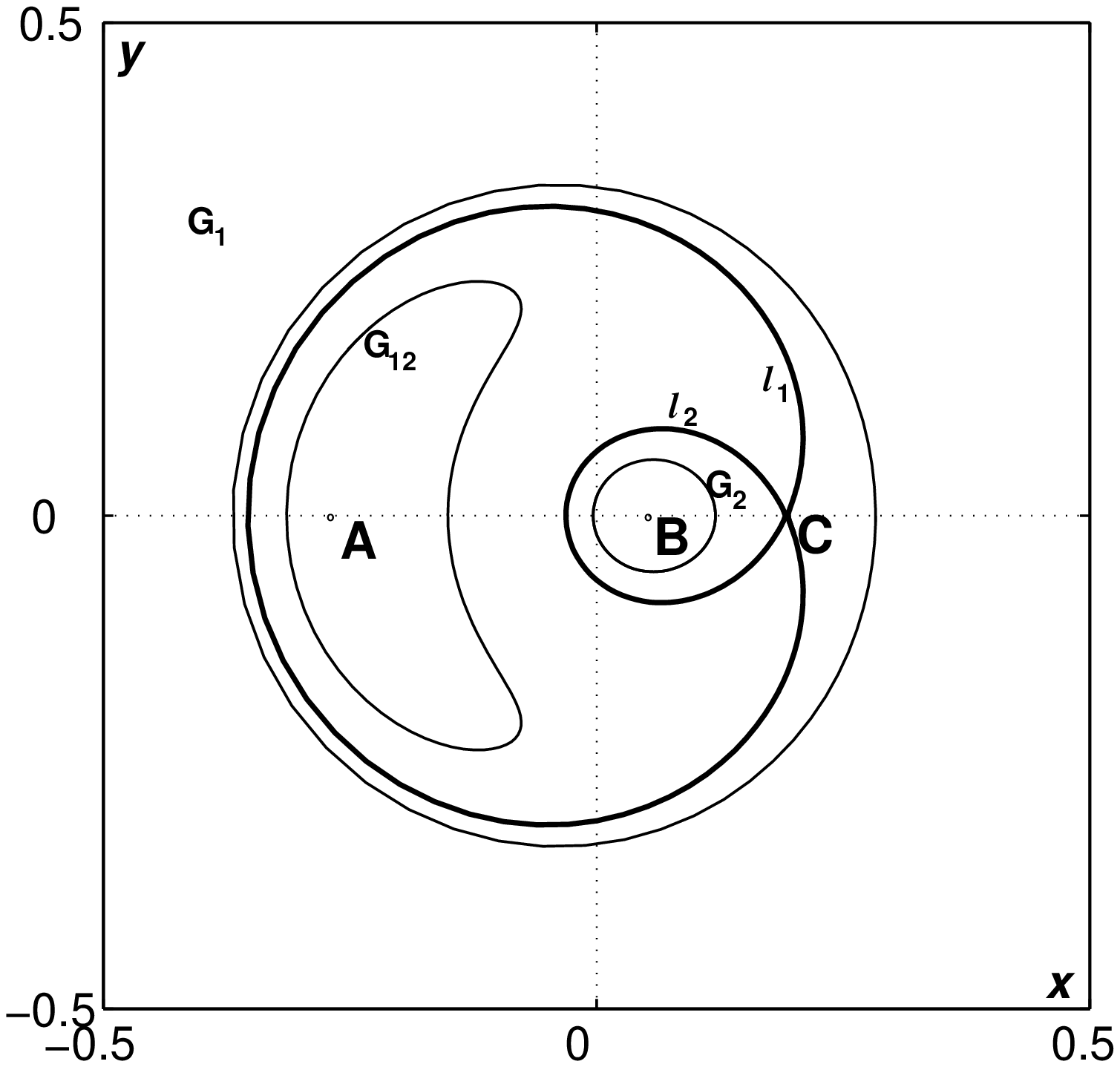,width=300pt}
\caption{\small{Phase portrait of the system at $\la = 0.1 >
\la_*$; $\tilde\mu = 0.01$.} } \label{portrait3}
\end{figure}

Phase portraits of (\ref{0.8}) at different constant values of $\la$ are presented in Figures \ref{portraits12} and \ref{portrait3}. If $\la < \la_* = \frac32 \mu^{2/3}$, there is one elliptic point A (Figure \ref{portraits12} a). At $\la > \la_*$, there are two elliptic points A and B, and one saddle point C (Figure \ref{portrait3}). In the latter case, the phase plane is divided by separatrices $l_1, l_2$ into three regions $G_1,G_{12},G_2$.

Let $H_c(\la)$ be the value of Hamiltonian (\ref{0.8}) at the saddle point C. Introduce $\calH = H - H_c$. Then we have $\calH <0$ in $G_{12}$, $\calH > 0$ in $G_1,G_2$, and $\calH = 0$ on the separatrices $l_1, l_2$.

As parameter $\la$ slowly grows with time,  curves $l_1, l_2$ slowly move on the phase plane. On time intervals of order $\mu^{2/3}\delta^{-1}$ their position and the areas of $G_{12}, G_2$ essentially changes. At the same time, the area surrounded by a closed phase trajectory at a frozen value of $\la$ is an adiabatic invariant of the system with slowly varying parameter $\la$ and is well preserved. Hence, phase points can cross $l_1, l_2$ leaving one of the regions $G_i$ and entering another region.

Assume that $\la = \la_*$ at $t=0$. The corresponding phase portrait is presented in Figure \ref{portraits12} b. The following assertion is valid: if a phase point is inside $l_1$ at $\la=\la_*$, it is captured in
$G_{12}$ at $t>0$ and stays there at least during time intervals of order $\delta^{-1}$. (More precisely, this may not be valid for phase points belonging initially to a narrow strip $-k_1 \delta^{6/5} \le \calH <0$, where $k_1$ is a positive constant.) Thus, all phase points initially (at $\la = \la_*$) inside $l_1$, except, maybe, for a narrow strip, are ``automatically'' captured into $G_{12}$. In \cite{Sincl}, this phenomenon was called ``automatic entry into libration''. The diameter of this domain is a value of order $\mu^{1/3}$.

A point captured in $G_{12}$ rotates around elliptic point A. As time grows, the point A on
the portrait slowly moves along $x$-axis in the negative direction. Therefore, the motion is a
composition of fast rotation around A and slow drift along $x$-axis. The area surrounded by each turn of the trajectory stays approximately the same (this area is an adiabatic invariant). Hence, the average distance between the phase point and the origin slowly grows,
corresponding to the growth of amplitude of the pendulum's oscillations in the original problem.

Consider now the case when a phase point is initially outside $l_1$. As $\la$ grows, the
area inside $l_1$ also grows, and at a certain moment the phase trajectory crosses $l_1$. This is due to the fact that the area surrounded by the phase trajectory is an adiabatic invariant, while the area $S_{12}$ of $G_{12}$ monotonously grows (see below (\ref{0.11})).  After crossing,  the phase point can continue its motion in $G_{12}$ during a time
interval of order at least $\delta^{-1}$ (capture into resonance) or can cross $l_2$
and continue its motion in $G_2$ (passage through the resonance without
capture). The area $S_2$ of $G_2$ also monotonously grows with time (see below (\ref{0.11})),
hence such a point cannot cross $l_2$ once more and return to $G_{12}$.

The scenario of motion after crossing $l_1$ strongly depends on initial conditions, and capture into $G_{12}$  can be considered as a random event. Its probability can be found according to the following formula:
\be
\mbox{Pr} = \frac{I_1 - I_2}{I_1},  \;\; \mbox{where} \; I_1(\la) = -\oint_{l_1} \frac{\pt \calH}{\pt
\la} \dd t, \;\; I_2(\la) = -\oint_{l_2} \frac{\pt \calH}{\pt \la} \dd t,
\label{0.9}
\ee
where the integrals are calculated at $\la = \La$, and $\La$ is the value of $\la$ at the time of crossing of $l_1$ found in the adiabatic approximation. One obtains:
\be
I_1(\Lambda) = \frac12 (2\pi - \Theta), \;\; I_2 = \frac{\Th}{2}, \;\; \Th = \arccos\left(
\frac{\Lambda}{2x_C^2} - 2 \right).
\label{0.10}
\ee
Here $\Th$ is the angle between the tangencies to $l_1$ at C, $0 \le \Th < \pi$, and $x_C$ is the $x$-coordinate of the saddle point C. Note that
\be
\frac{\dd S_2}{\dd\la} = I_2 \;, \;\;\; \frac{\dd S_{12}}{\dd\la} = I_1 - I_2.
\label{0.11}
\ee
Formula (\ref{0.9}) can be interpreted as follows. In a Hamiltonian system, phase
volume is conserved. As parameter $\la$ changes by $\Dt \la$, a phase volume $\Dt V_{1}$ enters
the region $G_{12}$. At the same time, a volume $\Dt V_2$ leaves this region and enters $G_2$. The
relative measure of points captured in $G_{12}$ is $(\Dt V_{1} - \Dt V_2)/\Dt V_2$. The integral
$I_1$ in (\ref{0.9}) is the flow of the phase volume across $l_1$, and $I_2$ is the flow across
$l_2$. Therefore, Pr gives the relative measure of points captured into $G_{12}$.

The results of this section were obtained assuming that the adiabaticity condition is valid. To express this condition in terms of the parameters of the system,  consider Hamiltonian
(\ref{0.8}). A typical scale of the corresponding phase portrait can be found using the condition
that the first and the last terms in the Hamiltonian are of the same order. Thus,
typical values of coordinates $x$ and $y$ on the portrait are of order of $\mu^{1/3}$. Hence, from (\ref{0.6}) and (\ref{0.5}) we find that
typical frequency of motion on this portrait is $\Om_{typ} \sim \dot \gm \sim \rho \sim \mu^{2/3}$. The adiabaticity condition implies that variation $\Dt\om$ of the driving frequency during a period of motion on the phase portrait is much smaller than the frequency of motion. Hence, it  can be written as $\Dt\om \sim \dt \Om_{typ}^{-1} \ll \Om_{typ}$ or $\dt \ll \Om_{typ}^2$. Thus we have $\dt \ll \eps^{4/3}$. At values of $\eps$ that do not satisfy this condition the adiabatic approximation does not work, and, in particular, capture into the resonance is impossible. In \cite{FajansFriedland:2001}, an expression was obtained for the threshold value of $\eps$ such that at smaller $\eps$ the capture into the resonance is not possible. Our estimate agrees with the  result of \cite{FajansFriedland:2001}.

Summarizing, we can say that at small enough initial amplitudes of oscillations (of order $\eps^{1/3}$ or less) the pendulum is necessarily captured into the 1:1 resonance with the external forcing of slowly decreasing frequency. If the initial amplitude is larger but still small, the capture occurs with probability of order 1 given by formula (\ref{0.9}). In the frames of model (\ref{0.2}), valid at small amplitudes, the captured pendulum cannot escape from the resonance. We study the system at larger values of the amplitude in the next section.

\section{Forced pendulum at a nonlinear resonance}\label{sec3}

Consider again Hamiltonian of the pendulum under the external forcing (\ref{0.1}). Introduce unperturbed Hamiltonian $H_0 = P^2/2 - \om_0^2 \cos Q$ and parameter
\be
\kappa^2 = \frac12\left(1 + \frac{h_0}{\om_0^2}\right),
\label{1.2}
\ee
where $h_0$ is a value of $H_0$. Thus, $0 \le \kp^2 < 1$ in the domain of oscillations of the pendulum, $\kp^2  = 1$ on the separatrix, and $\kp^2  > 1$ in the domain of rotations. To introduce the canonical action-angle variables $(I,\ffi)$, we note that in the domain of oscillations the exact solution for the unperturbed pendulum (i. e., at $\eps = 0$) has the form
\bea
P(t) = 2 \kp \om_0\, \cn (\om_0 t,\kp), \nonumber \\
Q(t) = 2 \, \arccos( \mbox{dn} (\om_0 t,\kp)),
\label{1.2a}
\eea
where $\cn$ and $\mbox{dn}$ are Jacobi elliptic functions (see, e. g., \cite{SagdeevZaslavsky}; the second formula can be obtained as a primitive of the first one). Thus, one can introduce canonical transformation  $(p,q) \mapsto(I,\ffi)$ with the following formulas
\bea
P = 2 \kp \om_0\, \cn (\frac{\om_0}{\Om}\ffi,\kp), \nonumber \\
Q = 2 \, \arccos( \mbox{dn} (\frac{\om_0}{\Om}\ffi,\kp)),
\label{1.2b}
\eea
where $\kp$ should be understood   as a function of $I$ defined with the  formula valid in the domain of oscillations of the pendulum (see, e.g., \cite{SagdeevZaslavsky}):
\be
I(h_0) = \frac{8\om_0}{\pi} \left[E(\kp) - (1-\kp^2)K(\kp) \right].
\label{1.16}
\ee
Here $E(\kp)$ and $K(\kp)$ are the complete elliptic integrals of the second and the first kind respectively:
\be
E(\kp) = \int^{\pi/2}_0 \sqrt{1-\kp^2\sin^2 u} \,\, \dd u, \,\,\,\,\, K(\kp) = \int^{\pi/2}_0 \frac{\dd u}{\sqrt{1-\kp^2\sin^2 u}}.
\label{1.17}
\ee
In (\ref{1.2b}), $\Om=\Om(\kp)$ is the frequency of oscillations of the unperturbed pendulum,
\be
\Om(\kp) = \frac{\pi}{2} \om_0 \frac{1}{K(\kp)}.
\label{1.17a}
\ee

Expanding the first equation of (\ref{1.2b}) into the Fourier series we obtain (see \cite{AbramowitzStegun}):
\be
P= 2 \om_0 \frac{2\pi}{K} \sum^\infty_{n=0} \frac{q^{n+1/2}}{1+q^{2n+1}} \cos\left[(2n+1)\ffi\right],
\label{1.4}
\ee
with
\be
q = \exp\left(-\frac{\pi K^{\prime}}{K}\right), \,\,\,\, K = K(\kp),  \,\,\,\, K^{\prime} = K(\kp'), \,\,\,\, \kp'^2 = 1-\kp^2.
\label{1.5}
\ee

To study dynamics near the 1:1 resonance between the pendulum's oscillations and the external forcing, introduce the so-called resonant phase $\gm = \ffi - \psi$ as a new variable.  We do this with a canonical change of variables $(I,\ffi) \mapsto (\tilde I, \gm)$ defined by generating function $W(\tilde I,\ffi) = \tilde I (\ffi - \psi)$. Thus we have
\be
I = \frac{\pt W}{\pt \ffi}= \tilde I, \,\,\,\, \gm= \frac{\pt W}{\pt\tilde I} =  \ffi - \psi,
\label{1.8}
\ee
and for the Hamiltonian expressed via the new variables we find
\be
\tilde H = H + \frac{\pt W}{\pt t} = H - I \om.
\label{1.9}
\ee
From now on, we omit tildes over $H$.

Fourier expansion for the coordinate $Q$ can be obtained as the primitive of the expansion for $P$ (see (\ref{1.4})) with substituting $\ffi = \gm +\psi$.   We plan to average the system near the resonance, hence we keep only the $n=0$ term in this expansion for $Q$:
\be
Q_r = 8 \frac{q^{1/2}}{1+q} \sin (\gm+\psi).
\label{1.6}
\ee
Substituting $ Q_r $ for $Q$ in (\ref{0.1}) and (\ref{1.9}) and averaging, we find that the Hamiltonian averaged near the resonance is
\be
H= H_0(I) + \eps A(I) \sin \gm - I \om, \,\,\,\, \mbox{where} \,\,\,\, A(I) = 4 \frac{q^{1/2}}{1+q}.
\label{1.7}
\ee
Here $H_0(I)$ is the unperturbed Hamiltonian of the pendulum in terms of the action variable $I$, and $A$ is also considered as a function of $I$. [Note, that $\kp$ is a monotonous function of $h_0$, and $h_0$ is a monotonous function of $I$. Here and below we write for brevity $I$ instead of $\kp(I)$ in the arguments of functions depending on $\kp$ ].

At the resonance $\dot \gm = 0$, hence $\dot \ffi = \om$. Therefore, the resonant value of the action variable $I_R$ is defined by the equation $\Om(I_R) = \om(\delta t)$, where $\Om(I) \equiv 2\pi/T$ is the frequency of oscillations of the unperturbed pendulum. Hence, $I_R$ is a function of the slow time $\tau \equiv \delta t$, i.e. $I_R = I_R(\tau)$.

Far from the resonance one can average over $\gm$ and obtain a system with conserved value of $I$. This corresponds to the fact that off-resonance perturbation does not change strongly the amplitude of the pendulum's oscillations.

\begin{figure}[h]
\center\epsfig{file=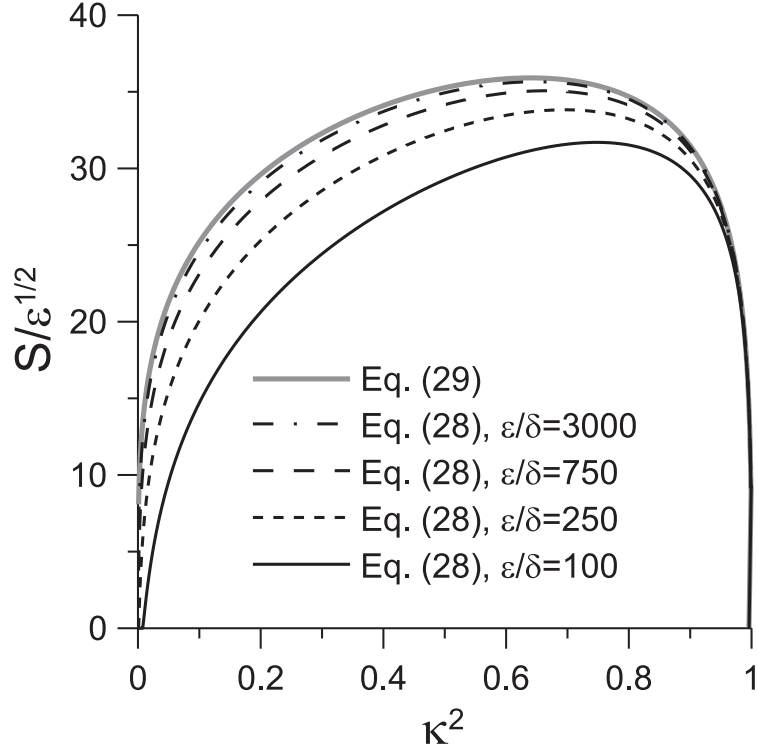, width=4in} \caption{\label{f1} Area (divided by $\sqrt{\eps}$) inside the separatrix on the phase portrait of the inner pendulum as a function of  $\kp^2$ at various values of the parameters.}
\label{figArea}
\end{figure}

Near the resonance, one can expand the Hamiltonian (\ref{1.7}) into series with respect to $(I-I_R)$. Retaining the main terms, we find:
\be
H = H_0 (I_R) + \left.\frac12 \frac{\pt^2 H_0}{\pt I^2}\right|_{I=I_R} (I-I_R)^2+  \eps A(I_R) \sin \gm.
\label{1.10}
\ee
Make a canonical change of variables $(I,\gm) \mapsto (\calP,\tilde\gm)$ defined with the generating function $W_1(I,\tilde\gm) = (I-I_R)\tilde\gm$. We find
\be
\calP= \frac{\pt W_1}{\pt \tilde\gm}= I-I_R, \,\,  \gm = \frac{\pt W_1}{\pt I} =\tilde\gm, \,\, \tilde H = H + \frac{\pt W_1}{\pt t}.
\label{1.11}
\ee
Omitting tildes, we find for the Hamiltonian
\bea
H=\La(\tau)+F(\calP,\gm), \,\,\, \mbox{where} \,\, \La(\tau) = H_0(I_R(\tau)), \nonumber \\
F = \frac12 g\calP^2 + d \sin \gm + \delta b \gm.
\label{1.12}
\eea
Here the coefficients $g,d,$ and $b$ are functions of $I_R(\tau)$ given by
\be
g= \left.\frac{\pt^2 H_0}{\pt I^2}\right|_{I=I_R}, \,\,\, d=\eps A(I_R), \,\,\, \delta b= -\frac{\pt I_R}{\pt t} = -\delta \frac{\pt I_R}{\pt \tau}.
\label{1.13}
\ee
The Hamiltonian $F$ is one of a pendulum under the action of the external torque. We shall call it ``the inner pendulum'' to distinguish it from the original pendulum (\ref{0.1}). Such Hamiltonians universally occur in resonant problems (see, e.g., \cite{Neishtadt:1999,NV:2006}). Coefficients $g,d,$ and $b$ are varying slowly, at a rate proportional to $\delta$, hence we can first consider the inner pendulum at frozen values of these coefficients. The phase portrait of this system can be one of the two types: if $|\delta b|< |d|$, there is a separatrix and the domain of oscillations on the portrait; if $|\delta b|\ge |d|$, there is no domain of oscillations. In the former case, the area $\calS$ inside the separatrix is given by the formula:
\be
\calS = 2\int_{\gm_m}^{\gm_s} \left(2 \frac{d}{g}\left[\sin \gm_s + \frac{\delta b}{d}\,\gm_s - \sin\gm- \frac{\delta b}{d}\,\gm\right]\right)^{1/2} \dd \gm,
\label{1.14}
\ee
where $\gm_s = 2\pi -\arccos(-\delta b/d)$ and $\gm_m$ is the root of equation $\sin\gm + \frac{\delta b}{d}\,\gm = \sin\gm_s + \frac{\delta b}{d}\,\gm_s$  satisfying $\gm_s-2\pi < \gm_m <\gm_s$. The area $\calS$ is presented in Figure \ref{figArea} at various values of the ratio $\eps/\dt$.

Phase portrait of the inner pendulum slowly evolves with time; in particular, the area $\calS$ is a function of the slow time $\tau$. The area surrounded by a phase trajectory inside the separatrix on the phase portrait of the inner pendulum is an adiabatic invariant (called the {\it inner adiabatic invariant}). Thus, while $\calS$ grows with time, such phase trajectories cannot leave the domain of oscillations of the inner pendulum.  On the other hand, additional phase volume appears inside the separatrix, and phase points can be captured into the domain of oscillations. In the captured motion, the value of $I$ strongly changes in such a way that, in the main approximation, the resonance condition $\Om(I) = \om(\tau)$ is preserved. As $\om(\tau)$ is a monotonously decreasing function, the amplitude of oscillations of the pendulum grows.

If $|\delta b|\ge |d|$, there is no separatrix on the phase portrait of the inner pendulum, and hence phase points cannot be captured into the resonance. Therefore at  given values of $\delta$ and initial $I$ there is a threshold value of the force amplitude $\eps_{th}$, such that if $\eps < \eps_{th}$, the capture is impossible (cf. \cite{Friedland:2001}). In particular, at small enough values of the initial action $I_0$ (and, correspondingly, small values of $\kp$) we have from (\ref{1.16}) and (\ref{1.7}) that $I_0 \approx 2\om_0\kp^2$ and $A \approx \kp \approx (2 \om_0)^{-1/2}\sqrt{I_0}$.  Hence, the condition $|\delta b|\ge |d|$ takes the form $\eps\sqrt{I_0}\le \dt |b| \sqrt{2\om_0}$. If $I_0\sim \eps^{2/3}$, we find $\eps_{th}^{4/3} \approx \dt |b| \sqrt{2\om_0}$. This agrees with the result of \cite{FajansFriedland:2001}, obtained for small oscillation amplitudes. Note, however, that at such relation between parameters $\eps$ and $\dt$ the system lacks adiabaticity, and, strictly speaking, the implemented approach is inadequate.

Consider now the important case $\delta \ll \eps$. One can neglect the  terms containing $\delta$ in the formula for $\calS$ (\ref{1.14}). The area inside the separatrix on the phase portrait of the inner pendulum is then given by the formula
\be
\calS = 16 \sqrt{-\frac{A\eps}{g}}
\label{1.15}
\ee
with $A$ and $g$ defined in (\ref{1.7}) and (\ref{1.13}), accordingly.

Our next object is to find an explicit expression for $g$. We have
\bea
g = \left.\frac{\pt^2 H_0}{\pt I^2}\right|_{I=I_R} = \left.\frac{\pt }{\pt h_0}\left(\frac{1}{\pt I/\pt h_0}\right)\frac{\pt h_0}{\pt I}\right|_{I=I_R} \nonumber \\
 = -\left.\frac{\pt }{\pt h_0}\left(\frac{1}{\Om}\right)\, \Om^3\right|_{I=I_R} = - \left. \frac{\pt}{\pt \kp}\left(\frac{1}{\Om}\right)\, \Om^3 \, \frac{\pt \kp}{\pt h_0}   \right|_{I=I_R} .
\label{1.18}
\eea
Using formulas for derivatives of elliptic integrals (see, e.g., \cite{AbramowitzStegun}), we find:
\be
g = \left.-\frac{\pi^2}{16 \kp^2 \kp'^2}\cdot \frac{E(\kp) - \kp'^2 K(\kp)}{[K(\kp)]^3}\right|_{I=I_R}.
\label{1.19}
\ee
In this expression, $g$ should be calculated at such a value of $h_0$ that $I(h_0) = I_R$. Substituting expression (\ref{1.7}) for $A$ at $I=I_R$ and (\ref{1.19}) into (\ref{1.15}), we obtain the expression for $\calS$ as a function of $\kp$. The plot of this function is presented in Figure \ref{figArea} (the uppermost curve). One can see that as the ratio $\eps/\dt$ grows, formula (\ref{1.15}) gives better and better approximation to (\ref{1.14}).

Let initially (at $t=0$) the pendulum have an oscillation amplitude corresponding to the value of action $I = I_0$ and the frequency $\Om = \Om_0$. Let the forcing frequency  $\om(\tau)$ at $t=0$ be larger than $\Om_0$. The frequency $\om$ slowly decreases with time, and at the time $\tau_*$ found from the equation $\om(\tau_*) = \Om(I_0)$, the pendulum is in the resonance with the forcing and $I_R = I_0$. Assume that $\calS = \calS_*$ at $\tau=\tau_*$. If $\pt \calS/\pt \kp>0$ at $I=I_0$, the pendulum can be captured into the resonance. The capture is a probabilistic phenomenon, and its probability Pr can be calculated (see \cite{Neishtadt:1975}). For the inner pendulum (\ref{1.12}) the formula for the probability of capture Pr at $\eps \ll 1$  is
\be
\mbox{Pr} = \frac{\calS^{\prime}}{2\pi |b|},
\label{1.20}
\ee
where prime denotes derivative with respect to $\tau$. To find $b$, we differentiate the resonance condition $\om(\tau) = \Om (I_R)$ and obtain $b = -I_R^{\prime} = -\om^{\prime}/g$. In the considered case $\eps \ll 1$ the probability Pr is a value of order  $\sqrt{\eps}$:
\be
\mbox{Pr} = \sqrt{\eps} \, \frac{8}{\pi} \, \frac{g}{\om^{\prime}}\, \left(\sqrt{- A/g}\right)^{\prime},
\label{1.21}
\ee
where all values should be calculated at $\tau = \tau_*$.

In the captured motion variable $I$ grows with time in such a way that the resonance condition $\om(\tau) = \Om(I)$ is preserved; accordingly, value of parameter $\kp$ also grows. The inner adiabatic invariant is approximately preserved. Thus, while $\calS$ grows with time, the captured phase point deepens more and more into the domain of oscillations of the inner pendulum. At a certain value of $\kp$ function $\calS(\kp)$ has a maximum (see Figure \ref{f1}). At larger values of $\kp$ the area $\calS$ is a monotonously decreasing function. Hence, as $I$ continues to grow with time, $\calS$ decreases. At a certain value of $I=I_{esc}$ the equality $\calS = \calS_*$ is again satisfied. As the inner adiabatic invariant is approximately preserved in the motion, at this moment the phase points captured at $\tau = \tau_*$  cross the separatrix of the inner pendulum and leave the domain of oscillations. Accordingly, the amplitude of the [original] pendulum's oscillations stops growing and the pendulum escapes from the resonance. Note that, therefore, the time when the escape occurs is determined by the initial amplitude of the pendulum's oscillations.

Assume that the pendulum was captured into the resonance at a small initial value of action $I=I_0 \sim \eps^{2/3}$. Thus, when the capture occurs, the area of the domain $G_{12}$ (see Figure \ref{portrait3}) is $\calS_* \sim \mu^{2/3} \sim \eps^{2/3}$ (c.f. Section \ref{sec2}). The escape from the resonance occurs at  $I=I_{esc}$ (or $\kp=\kp_{esc}$), and  $\calS(\kp_{esc}) = \calS_*$. Obviously, the value $\kp_{esc}$ is close to 1: $0<1-\kp_{esc}\ll 1$. Using formulas for asymptotics of the elliptic integrals at $\kp \approx 1$ (see, e. g., \cite{AbramowitzStegun}) we find from (\ref{1.7}), (\ref{1.19}), and (\ref{1.15}) that
\be
g \approx \frac{\pi^2}{4(1-\kp)[\ln(1-\kp)]^3}, \,\,\, A \approx 2,
\label{1.22a}
\ee
and thus
\be
\calS(\kp_{esc}) \approx \frac{128}{\pi} \sqrt{\eps} \, \sqrt{1-\kp_{esc}}\, |\ln\sqrt{1-\kp_{esc}}|^{3/2}.
\label{1.22}
\ee
Hence, we have the estimate
\be
1-\kp_{esc} \approx\frac{\pi^2}{16384} \frac{ \calS_*^2}{\eps|\ln(\calS_*/\sqrt{\eps})|^3} \sim \eps^{1/3}|\ln\eps|^{-3}.
\label{1.23}
\ee

Now consider the case of small $I_0$ using formula (\ref{1.15}) for $\calS$.   From (\ref{1.16}) we find that at small enough values of $I_0$ one has $I_0 \approx 2\om_0\kp^2$. If the capture into the resonance occurs at $I=I_0$, one finds from (\ref{1.15}) that  $\calS_* \approx 32 \sqrt{2\eps\kp} \approx 32 (2/\om_0)^{1/4}\sqrt{\eps} I_0^{1/4}$. (Note, that if $I_0 \sim \eps^{2/3}$, this formula for $\calS_*$ agrees the estimate given in the previous paragraph.) Substituting this expression for $\calS_*$ into equation $\calS(\kp_{esc}) = \calS_*$ and using (\ref{1.22}), we find
\be
1-\kp_{esc} \approx 4 \pi^2\sqrt{2/\om_0} \, \sqrt{I_0}|\ln I_0|^{-3}.
\label{1.24}
\ee

It is interesting to compare  the value of the energy $h_{esc}$ at the escape from the resonance with the width of the stochastic layer surrounding the separatrix of the pendulum (see, e. g., \cite{AKN,SagdeevZaslavsky}).    An estimate of this width can be obtained as follows. We assume that escape from the resonance occurs close to the separatrix of the pendulum. Consider the equation of motion of the pendulum
\be
\ddot Q + \om_0^2 \sin Q = -\eps \cos \psi
\label{1.25}
\ee
with fixed value of $\dot \psi = \om \sim |\ln (h_s-h)|^{-1}$, corresponding to the motion close to the separatrix (here $h$ is the energy of the unperturbed pendulum, and  $h_s = \om_0^2$ is its energy on the separatrix). The width of stochastic layer in this problem was estimated earlier (see \cite{SagdeevZaslavsky}, Section 3 of Chapter 5). A phase point is in the stochastic layer provided that $|h-h_s| \preceq \eps \om$. Hence, in the stochastic layer $|h-h_s| \preceq \eps |\ln \eps|^{-1}$.
Comparing this with the estimate in (\ref{1.23}), we see that at small enough $\eps$  even in the case of  ``automatic'' capture into the resonance the escape can occur well before the captured trajectory enters the stochastic layer.

\begin{figure}
\center\epsfig{file=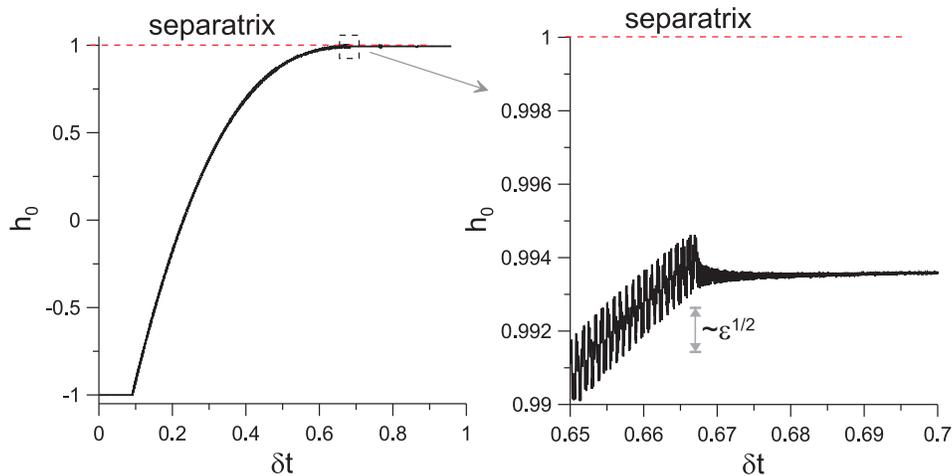, width=5in}
\caption{Energy $h_0$ as a function of time along a phase trajectory with the  automatic capture. Right panel shows a fragment of the trajectory in the vicinity of the separatrix near the escape from the resonance. Scale $\sim \varepsilon^{1/2}$ is indicated in the right panel. System parameters are $\varepsilon=10^{-6}$, $\delta=\varepsilon/750$.}
\label{fig1}
\end{figure}

To check  our conclusions  we perform numerical integration of the initial system (\ref{0.1}). We take frequency $\omega_0=1$ and frequency $\omega=1.1(1-\delta t)$. Phase trajectories start at the initial moment of time $t=0$, and we calculate variation of the energy $h_0$ along a trajectory while $\delta t\leq 1$. Results of numerical integration for a phase trajectory with initial conditions in the domain  of the automatic capture are presented in Figure \ref{fig1}. At  $\delta t \approx 0.1$ the phase point becomes captured and the energy $h_0$ starts growing. In the vicinity of the separatrix ($h_0=1$) the phase point escapes from the resonance. The right panel of Fig. \ref{fig1} shows that the escape occurs well before the separatrix and even before the trajectory reaches the stochastic layer. After the escape the  energy stays constant. The  amplitude of the energy oscillations in the captured state is  $\sim \varepsilon^{1/2}$.

\begin{figure}
\center\epsfig{file=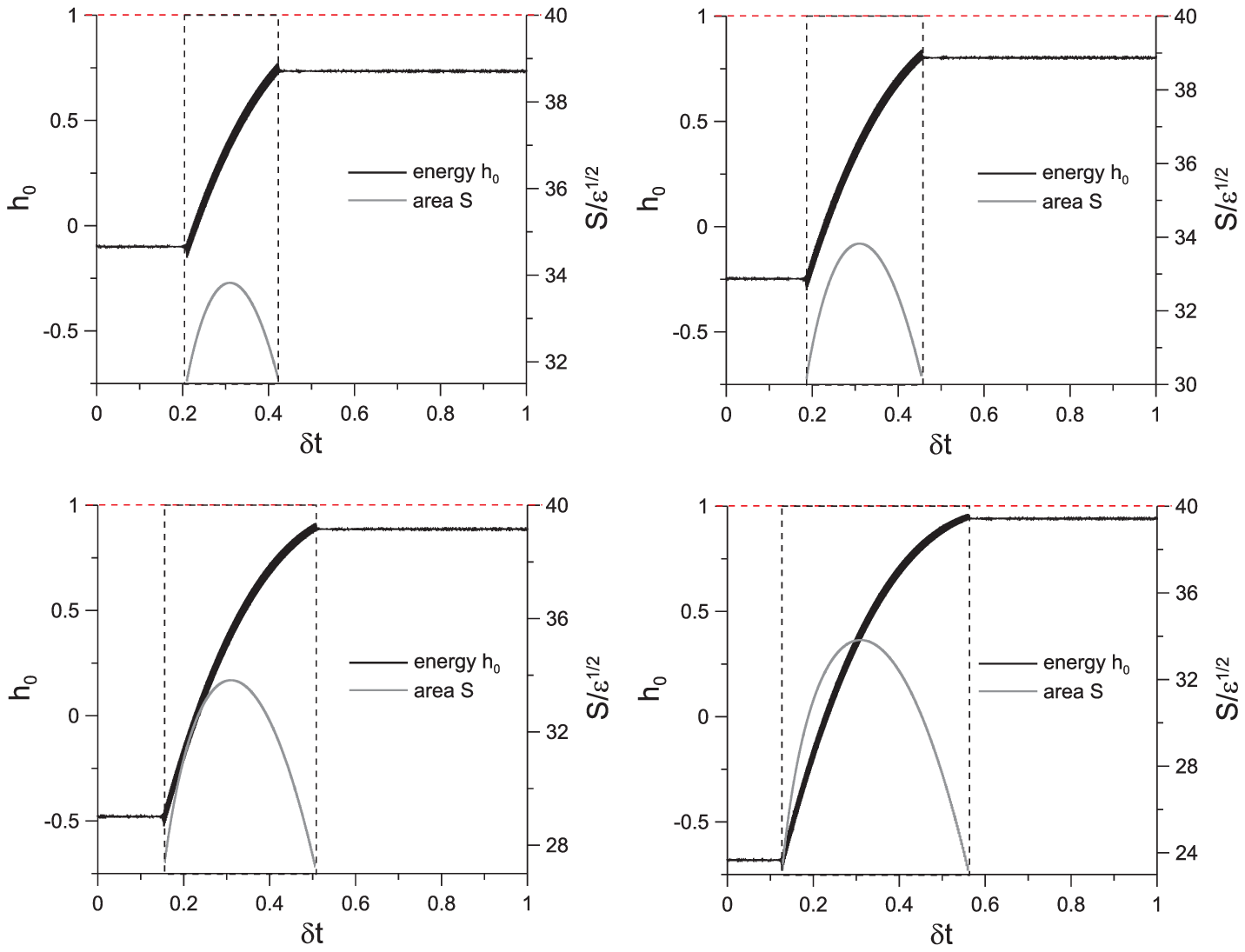, width=5in}
\caption{Energy $h_0$ as function of time for phase trajectories without automatic capture. Four different values of initial energy are considered. Grey curves show the area surrounded by the separatrix of the inner pendulum in the resonance.  System parameters are $\varepsilon=10^{-4}$, $\delta=\varepsilon/250$.}
\label{fig2}
\end{figure}

Figure \ref{fig2} shows four phase trajectories  without automatic capture, i.e. their initial energies  are chosen so that the probability of capture is less than one. We also calculate the area surrounded by the separatrix of the inner pendulum (see (\ref{1.14})) and show its time evolution while the phase point is captured. One can see that the captures occur when $S$ grows, while escapes from the resonance occur when it decreases. In each case, the escape from the resonance occurs at the  value of $S$ equal to its value  at the instant of the capture.

\section*{Acknowledgements}

The work was supported in part by the Russian Foundation for Basic Research (project no. 13-01-00251) and Russian Federation Presidential Program for the State Support of Leading Scientific Schools (project NSh-2519.2012.1). Work of A.V.A. and V.A.A. was also partially supported by Russian
Academy of Science (OFN-15).

\end{document}